\documentclass[fleqn,usenatbib,nofootinbib]{mnras}
\usepackage{graphicx}
\usepackage{amsmath,amssymb}
\usepackage{lastpage}
\usepackage[all]{hypcap}
\usepackage[T1]{fontenc}
\usepackage{float}
\usepackage{subfigure}
\usepackage{afterpage}
\usepackage[usenames]{color}
\usepackage{mathrsfs}
\usepackage{upgreek}
\usepackage{hyperref}
\usepackage{gensymb}
\usepackage{color}
\usepackage[dvipsnames]{xcolor}

\newcommand{\cs}{\texttt{CSiBORG}}
\newcommand{\csi}{\texttt{CSiBORG}$^{-1}$}
\newcommand{\reff}{R_\text{eff}}

\title{Catalogues of voids as antihalos in the local Universe}
\author[H.~Desmond et al.]{
Harry~Desmond$^{1}$\thanks{E-mail: harry.desmond@physics.ox.ac.uk}, Maxwell L. Hutt$^1$, Julien Devriendt$^1$ and Adrianne Slyz$^1$
\\
$^{1}$Astrophysics, University of Oxford, Denys Wilkinson Building, Keble Road, Oxford OX1 3RH, UK\\
}

\pubyear{2021}

\begin{document}
\label{FirstPage}
\pagerange{\pageref{FirstPage}--\pageref{LastPage}}
\maketitle

\begin{abstract}
A recently-proposed algorithm identifies voids in simulations as the regions associated with halos when the initial overdensity field is negated. We apply this method to the real Universe by running a suite of constrained simulations of the 2M++ volume with initial conditions inferred by the \texttt{BORG} algorithm, along with the corresponding inverted set. Our 101 inverted and uninverted simulations, spanning the \texttt{BORG} posterior, each identify $\sim$150,000 ``voids as antihalos'' with mass exceeding $4.38\times10^{11} \: \mathrm{M_\odot}$ (100 particles) at $z=0$ in a full-sky sphere of radius 155 Mpc$/h$ around the Milky Way. We calculate the size function, volume filling fraction, ellipticity, central density, specific angular momentum, clustering and stacked density profile of the voids, and cross-correlate them with those produced by \texttt{VIDE} on the same simulations. We make our antihalo and \texttt{VIDE} catalogues publicly available.
\end{abstract}


\section{Introduction}
\label{sec:intro}

% General importance of voids
% Different voidfinding methods
% Recap of antihalo method literature and the advantages of that
% Using that to make a void catalogue of the real universe

Cosmic voids are the lowest density regions of the Universe, occupying most of its volume. While most effort to constrain structure formation and cosmology has focused on the regions of greatest density---galaxies and halos---voids are emerging as powerful probes of the constituents and physics of the Universe (see \citealt{white_paper} and references therein). A large part of their constraining power stems from the fact that they evolve linearly for far longer than filaments, walls and halos, making them more amenable to analytic modelling. However, traditional watershed voidfinders
%such as \texttt{VIDE} \citep{Zobov}
do not identify purely linearly evolving regions, making their products difficult to connect to theoretical expectations \citep{Nadathur_Hotchkiss}. 
%Added to this is the traditional requirement that the density field used for voidfinding be estimated from the observed distribution of \emph{galaxies}, which are sparse and biased tracers of the underlying dark matter.
In addition, voids are normally estimated from the observed distribution of \emph{galaxies}, which are sparse and biased tracers of the underlying dark matter.

Recently, \citet{Pontzen_1, Pontzen} and~\citet{AntiClusters} have developed a new voidfinding algorithm for $N$-body simulations in which voids in one simulation are comprised of particles belonging to halos in a corresponding \emph{inverted} simulation where the initial overdensity, displacement and velocity fields are negated. This more physically-motivated, dynamical definition yields voids with sensible 1-point statistics and density profiles, and which furthermore are well described by the Zel'dovich approximation down to $M=10^{13} \: \mathrm{M_\odot}/h$. However, until now this voidfinder has not been applied to real data due to the requirement that the simulations on which it operates be \emph{constrained} to match the local Universe. Here we produce a suite of catalogues of ``voids as antihalos'' out to $\sim$155 Mpc/$h$ from the Milky Way by running $N$-body simulations with the phases of the initial density modes tuned to match the observed large-scale structure in the 2M++ catalogue \citep{2Mpp}. In addition to securing the benefits of the antihalo definition, this suite allows full marginalisation over the uncertainties in the initial conditions for Bayesian post-processing applications. We make it publicly available.\footnote{\url{https://zenodo.org/record/5503610} \citep{zenodo}}

Sec.~\ref{sec:method} describes our simulations and methods for making the void catalogues. Sec.~\ref{sec:results} displays the properties of the voids, including 1- and 2-point functions and stacked density profiles. Sec.~\ref{sec:disc_conc} summarises, discusses remaining open questions for the antihalo voidfinder, and proposes applications.

\section{Method}
\label{sec:method}

% BORG -> CSiBORG
% Inverting the ICs
% Voronoi tessellation
% Identifying the voids
% Calculating the void properties, e.g. stacked density, axis ratios, specific AM
% Cross-correlations etc
% Repeating over the full set for uncertainties

\subsection{\texttt{BORG} -- \cs\ -- \csi}
\label{sec:BORG}

Our method begins with the \texttt{BORG} algorithm, a probabilistic Bayesian inference machine for forward modelling the galaxy number density field and comparing to observations to constrain the phases of the dark matter density modes along with nuisance parameters including galaxy bias \citep{BORG_1, BORG_2}. We use the 2M++ reconstruction \citep{BORG_2Mpp}, which employs a box of side length 677.77 Mpc/$h$ (centred on the Milky Way) split into $256^3$ voxels, and follows the evolution of the density field from $z=69$ to the present using a particle-mesh gravity solver \citep{BORG_PM}. The parameters are inferred with a Hamiltonian Monte Carlo sampler, producing a chain of $\sim$100 autocorrelation lengths in the voxel densities. Only in the region in which the 2M++ survey has high completeness (within $\sim$155 Mpc$/h$), however, is the density field well constrained.

We use the initial conditions (ICs) in the converged part of the \texttt{BORG} chain to run a suite of high-resolution N-body simulations, dubbed \texttt{CSiBORG} (``Constrained Simulations in BORG''), using the \texttt{RAMSES} adaptive mesh refinement code \citep{Ramses}. These simulations perform a zoom-in on a sphere of radius 155 Mpc/$h$ around the Milky Way within which the \texttt{BORG} ICs are augmented with white noise to a resolution of $2048^3$, giving a minimum particle mass $m_p = 4.38\times10^9 \: \mathrm{M_\odot}$. Refinement is performed only in this central sphere, although the full \texttt{BORG} box is retained for accurate modelling of longer wavelength modes. To prevent numerical artifacts at the edge of the zoom-in region, a spherical buffer region of width $\sim10$ Mpc/$h$ is placed around the sphere in which the resolution degrades gradually to the background \texttt{BORG} value. \texttt{CSiBORG} contains 101 simulations in all with ICs spanning the full range of the \texttt{BORG} posterior, allowing the uncertainties in the \texttt{BORG} inference to be marginalised over in post-processing. Halos in the high-resolution region are found on the fly using the \texttt{mergertree} patch to \texttt{RAMSES}, which implements the watershed halofinder \texttt{PHEW} \citep{PHEW}. \texttt{CSiBORG} is a general-purpose constrained simulation suite and has previously been used in \citet{Bartlett}. \texttt{BORG} and \texttt{CSiBORG} use equatorial coordinates and the cosmology $T_\text{CMB} = 2.728$ K, $\Omega_m = 0.307$, $\Omega_\Lambda = 0.693$, $\Omega_b = 0.04825$, $H_0 = 70.5$ km s$^{-1}$ Mpc$^{-1}$, $\sigma_8 = 0.8288$ and $n = 0.9611$.

We have now run a corresponding set of simulations---\texttt{CSiBORG}$^{-1}$---in which the overdensities, particle displacements and velocities are the negative of their \cs\ counterparts. We modify \texttt{RAMSES} so that particles with the same undisplaced positions in \cs\ and \csi\ are assigned the same IDs. 
%particles with the same \emph{undisplaced} positions are assigned the same IDs, allowing us to identify void particles in \cs\ as those belonging to halos in \csi.
This provides 101 realisations of the local Universe in which particles are assigned both halo and antihalo flags.
%This adds value to the \cs\ suite: we now have 101 high-resolution realisations out to 155 Mpc/$h$ in which particles are assigned both halo and antihalo flags.
Here we work solely with the $z=0$ snapshots. Voids have previously been identified by applying \texttt{VIDE} to the SDSS \texttt{BORG} reconstruction in \citet{Florent_1}.

\subsection{Calculating void properties}
\label{sec:void_calcs}

To identify the antihalos in \cs\ we perform a Voronoi decomposition using \texttt{Voro++} \citep{Voro}. We first cut a cube out of each simulation just wide enough to enclose the full buffer region and hence containing all particles of mass less than the maximum. Although we are ultimately interested only in the $4.41\times10^8$ lowest-mass particles, including their neighbours is necessary for the Voronoi cells to be computed correctly at the edge of the zoom-in region.
%We perform the decomposition with \texttt{Voro++} \citep{Voro}.

A void in \cs\ comprises the set of particles indexed to a particular halo in \csi. Although \texttt{PHEW} is able to identify substructure, here we consider only main halos in \csi, mapping each particle to the highest-level (most massive) halo to which it belongs. We exclude voids with fewer than 100 particles
%($M<4.38\times10^{11} \: \mathrm{M_\odot}$)
which may be imperfectly resolved and unlikely correspond to true underdensities. For each of the remaining voids we calculate the following properties:

\begin{itemize}

\item $\vec{x}_\text{VWB}$: The position of the volume-weighted barycentre (VWB) $\sum_k \vec{x}_k V_k / \sum_k V_k$, where $V_k$ is the Voronoi volume of particle $k$ belonging to the void.
%By taking account of the positions of all the void particles, this is less sensitive to shot noise than the position of the lowest-density particle and proVIDEs more information about the void's boundaries.

\item $M$: The summed mass of the particles in the void. By construction this equals the halo mass in \csi.
%so that the halos and antihalo mass functions are identical.

\item $V$: The summed Voronoi volumes of the void particles.

\item $\rho_0$: The central density, defined as the average volume-weighted density within the sphere around the VWB that encloses 64 particles: $64 \: m_p / \sum_{k=1}^{64} V_k$.

\item $X, Y$: The two 3D ellipsoidal axis ratios. These are ratios of the square roots of the eigenvalues of the inertia tensor $I_{ij} \equiv \sum_k (x_{i,k} - x_{\text{VWB},i,k}) (x_{j,k} - x_{\text{VWB},j,k}) / N_\text{part}$, where $N_\text{part}$ is the number of particles in the void. 
%The principal axis lengths are the square roots of the eigenvalues of $I$.

% \item $\vec{j}$: The specific angular momentum of the void in the three Cartesian directions within the box: $j_i = \sum_k (\vec{x}_k - \vec{x}_{\text{VWB},k}) \times (\vec{v}_k - \vec{v}_{\text{VWB},k}) / N_\text{part}$, where $\vec{v}_\text{VWB} = \sum_k \vec{v}_k V_k / \sum_k V_k$.

\end{itemize}

% \begin{table}
%   \begin{center}
%     \begin{tabular}{l|r|r|r|}
%       \hline
% 					Statistic & \texttt{sparc}	        & Model mean  		& Discrepancy/$\sigma$\\ 
%       \hline
% \rule{0pt}{3ex}
%       $\rho$	& $-0.20$	& $0.00$         	& $2.2$\\
% \rule{0pt}{3ex}
%       $s_\text{BTFR}$ (dex)           	& $0.029$	& $0.064$        	& $3.6$\\
% \rule{0pt}{3ex}
%       \textquotedbl ($M_\text{b}>10^{9.5} M_\odot$)           	& $0.027$	& $0.053$        	& $2.1$\\
% \rule{0pt}{3ex}
%       $N_f$             		& $27$ 		& $33.4$         	& $2.2$\\
% \rule{0pt}{3ex}
%       $q$                 	& $0.003$	& $0.039$          	& $3.0$\\
%       \hline
%     \end{tabular}
%   \caption{Comparison of \texttt{sparc} and AM BTFR statistics. The 3rd row shows $s_\text{BTFR}$ for $M_\text{b} > 10^{9.5} M_\odot$ galaxies only.}
%   \label{tab:table}
%   \end{center}
% \end{table}

From these we can calculate the effective radius, $R_\text{eff} \equiv \left(\frac{3 V}{4 \pi}\right)^{1/3}$, and the lengths of the three principal axes defined so that their product is $\reff^3$: $R_x \equiv R_\text{eff} \: (XY)^{-1/3}$, $R_y \equiv R_\text{eff} \: (X^2/Y)^{1/3}$, $R_z \equiv R_\text{eff} \: (Y^2/X)^{1/3}$. We find the ellipsoidal approximation to be good for the majority of antihalos, although $\sim$15\% have disconnected regions and/or strings of particles extending from their centres.
%defined so that $R_x R_y R_z = \reff^3$.

We also calculate the density profiles of the voids in stacks. We compare four methods for this, two of which average in spherical annuli (scaled by $R_\text{eff}$) around the VWB, and the other two in ellipsoidal annuli aligned with the principal axes. In practice, the latter calculates
\begin{equation*}
d_{\text{ell}, k} = \left(\left(\frac{x'_k-x'_\text{VWB}}{R_x}\right)^2 + \left(\frac{y'_k-y'_\text{VWB}}{R_y}\right)^2 + \left(\frac{z'_k-z'_\text{VWB}}{R_z}\right)^2\right)^\frac{1}{2}
\end{equation*}
% \begin{align} 
% &d_k = \left(\left(\frac{x_k-x_\text{VWB}}{\reff}\right)^2 + \left(\frac{y_k-y_\text{VWB}}{\reff}\right)^2 + \left(\frac{z_k-z_\text{VWB}}{\reff}\right)^2\right)^{1/2} \\
% &d_{\text{ell}, k} = \left(\left(\frac{x'_k-x'_\text{VWB}}{R_x}\right)^2 + \left(\frac{y'_k-y'_\text{VWB}}{R_y}\right)^2 + \left(\frac{z'_k-z'_\text{VWB}}{R_z}\right)^2\right)^{1/2} \nonumber
% \end{align}
for each particle (not only those belonging to the void) with $d_{\text{ell}, k}<5$, where $\{x', y', z'\}$ is a rotated coordinate system aligned with the principal axes. For both types of averaging we consider two estimators for the stacked density profile. The first is the volume-weighted average density in each annulus with a small correction to remove bias in the estimate of the Poisson mean \citep{Nadathur}:
\begin{equation}\label{eq:dens2}
\bar{\rho}_j = m_p \frac{\left(\sum_i N_{ij}\right) + 1}{\sum_i V_{\text{shell},ij}},
\end{equation}
where $N_{ij}$ is the number of particles in shell $j$ of void $i$ and $V_{\text{shell},ij}$ is the volume of the $j$th spherical or ellipsoidal shell of void $i$. The second uses the Voronoi volumes of the particles instead of the volumes of the annuli:
\begin{equation}\label{eq:dens3}
\bar{\rho}_j = m_p \frac{\sum_i N_{ij}}{\sum_i \sum_{k=1}^{N_{ij}} V_k}.
\end{equation}
Eq.~\ref{eq:dens3} is more robust to boundary effects than Eq.~\ref{eq:dens2} 
%as well as tying in more naturally with the Voronoi tessellation that has already been performed to calculate particle volumes,
and is therefore to be preferred \citep{Nadathur}.
% We do not consider their ``naive'' method which is prone to edge effects and statistically biased.
% We stack separately in two mass bins: one below $10^{13} M_\odot/h$ and one above.
We only include in the stack voids with $M>10^{13} \: \mathrm{M_\odot}/h$.

\subsection{Comparison with \texttt{VIDE} and cross-correlations}
\label{sec:VIDE}

% How we run VIDE
% Dependence on subsample density
% Cross correlations with antihalos & halos

For a first comparison of our antihalo catalogues with catalogues of voids identified by other means, we run \texttt{VIDE} \citep{VIDE} on \cs. This uses \texttt{ZOBOV} \citep{Zobov} to identify voids via a hierarchical watershed procedure, beginning with a particle-by-particle density estimate through a Voronoi tessellation. 
%exactly as described above for antihalos.
We randomly sub-sample the \cs\ particles to a density of 0.2 (Mpc$/h$)$^{-3}$ before performing the tessellation, and require $\reff>2$ Mpc$/h$ to remove voids smaller than the mean particle separation. We use the default density cut $\bar{\rho}(R<\reff/4) \le 0.2\:\rho_m$ and merging threshold $0.2\:\bar{n}$. This limits the hierarchy of \texttt{ZOBOV} such that there are many top-level voids, and only these ``parent'' voids are included in our analysis. We also run \texttt{VIDE} on one realisation from \cs\ with a higher (2 (Mpc$/h$)$^{-3}$) and lower ($4\times10^{-3}$ (Mpc$/h$)$^{-3}$) sub-sampling density.

We cross-correlate antihalos, \texttt{VIDE} voids and halos in \cs\ and \csi\ in the zoom-in region between 5 and 35 Mpc$/h$ using \texttt{Corrfunc} \citep{corrfunc}.
%\mh{Auto-correlation functions worth including?}

\section{Results}
\label{sec:results}

% Distributions of the various quantities + VSF, VMF etc
% Positions of voids
% Cross-correlations, comparison with VIDE
% Stacked density profiles by the various methods

\subsection{1-point functions of void statistics}
\label{sec:1-pt}

% We begin by showing the distributions of the calculated void statistics. In each case we average over the 100 \cs\ realisations, although variations between the realisations in these distributions are very small. The standard deviation is shown explicitly for the size and volume fraction functions. The results in this section may be trivially reproduced from the public void catalogues.

\begin{figure}
  \centering
  \includegraphics[width=0.5\textwidth]{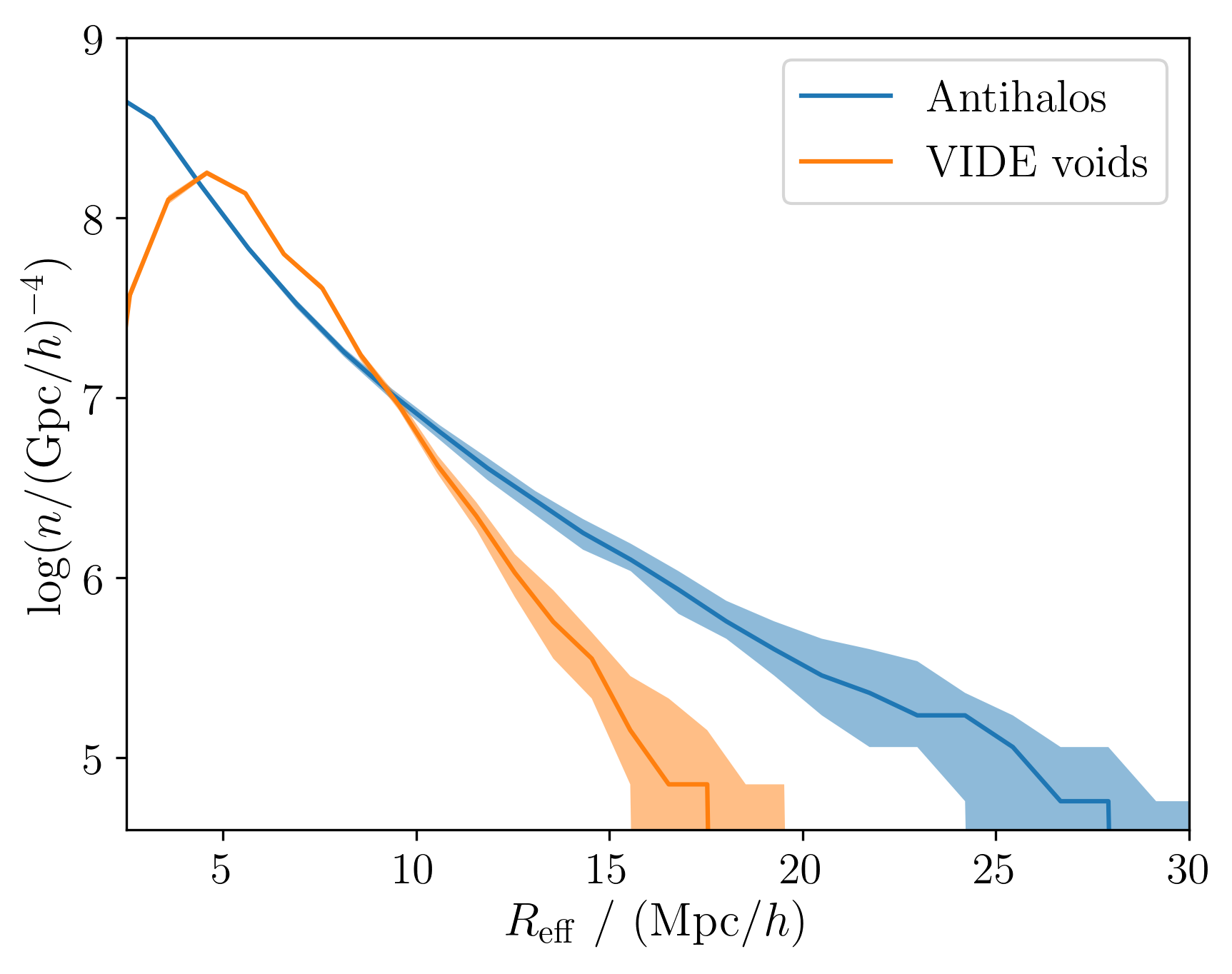}
  \caption{The differential size function of antihalos and \texttt{VIDE} voids. The solid line shows the median over all \cs\ realisations and the shaded band the $68\%$ confidence region.}
  \label{fig:VSF_comb}
\end{figure}

Fig.~\ref{fig:VSF_comb} shows the median differential void size function (VSF) across all 101 realisations and the $68\%$ confidence region between them, along with the corresponding result from \texttt{VIDE}. To enhance compatibility with \texttt{VIDE}, in all plots we restrict to antihalos with $\rho_0\le0.2\:\rho_m$. We only include objects with centres within $150$ Mpc/$h$ as those further out may be impacted by the buffer region. We see that our method produces both smaller and larger voids than \texttt{VIDE}. The VSF of \texttt{VIDE} is a strong function of the tracer density, with larger $\bar{n}$ producing more small voids and fewer large ones;
%As described in \citet{Sutter}, this occurs in watershed voidfinders because decreasing tracer density thins the walls and filaments at the boundaries between voids, eventually leading to their merger into larger structures.
with our default sub-sampling density of 0.2 (Mpc$/h$)$^{-3}$ there are $>$2$\times$ fewer \texttt{VIDE} voids than antihalos, although there are $>$2$\times$ more when $\bar{n} = 2$ (Mpc$/h$)$^{-3}$. We find a similar \texttt{VIDE} and antihalo VSF at the large radius end when $\bar{n} = 4\times10^{-3}$ (Mpc$/h$)$^{-3}$, although there are only $\sim$250 \texttt{VIDE} voids in total in that case.
%many fewer small voids are produced in that case such that there are only $\sim$250 in total.
%with the largest voids being roughly 30 Mpc$/h$ in radius. This is the same sub-sampling density as used by \citet{Sutter}, with whom our VSF agrees closely. However, only $\sim$500 voids are produced within 155 Mpc$/h$ at this density.

These voids are significantly smaller than those produced by conventional voidfinders applied to the galaxy field due to the higher tracer density afforded by the simulations. This is useful to boost statistics and access smaller-scale information. We verify the strong positive correlation between $R_\text{eff}$ and the mass of antihalos identified in \citet{AntiClusters}, which becomes almost monotonic for $M \gtrsim 10^{15} \: \mathrm{M_\odot}/h$. This is in stark contrast to \texttt{VIDE} where the correlation is weak.

\begin{figure}
  \centering
  \includegraphics[width=0.5\textwidth]{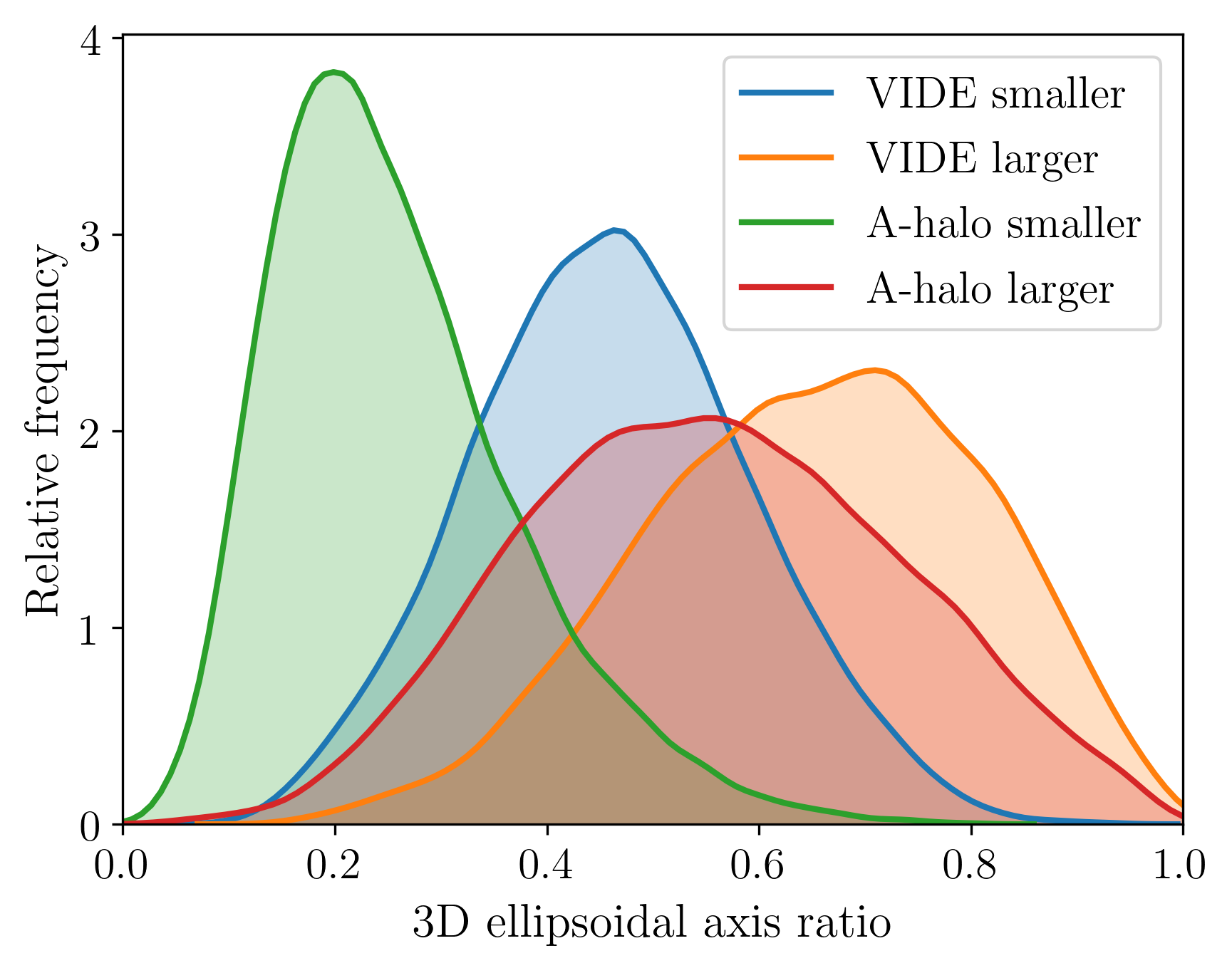}
  \caption{The distribution of ellipsoidal axis ratios of \texttt{VIDE} voids and antihalos in realisation \texttt{8740}.}
  \label{fig:axes}
\end{figure}

A reason to be wary of naive interpretation of $R_\text{eff}$ is given in Fig.~\ref{fig:axes}, which shows the antihalos to be quite strongly elliptical.
%and significantly more so than the \texttt{VIDE} voids.
The modal ratio of principal axis lengths is $\sim$1:2:5. The voids are more elliptical than the corresponding halos because they are not bound structures supported by self-gravity, but rather prone to deformation by the high-density filaments and sheets that surround them. This has an important impact on their stacked density profiles (Sec.~\ref{sec:dens}). We have checked that the antihalo major axes do not preferentially align with the line of sight, indicating that redshift space distortions are properly accounted for in \texttt{BORG} \citep{BORG_PM}. Figs.~\ref{fig:axes} and \ref{fig:stacked_density} use a randomly-chosen step number of the \texttt{BORG} chain (\texttt{8740}), but all realisations produce consistent results.

\begin{figure}
  \centering
  \includegraphics[width=0.5\textwidth]{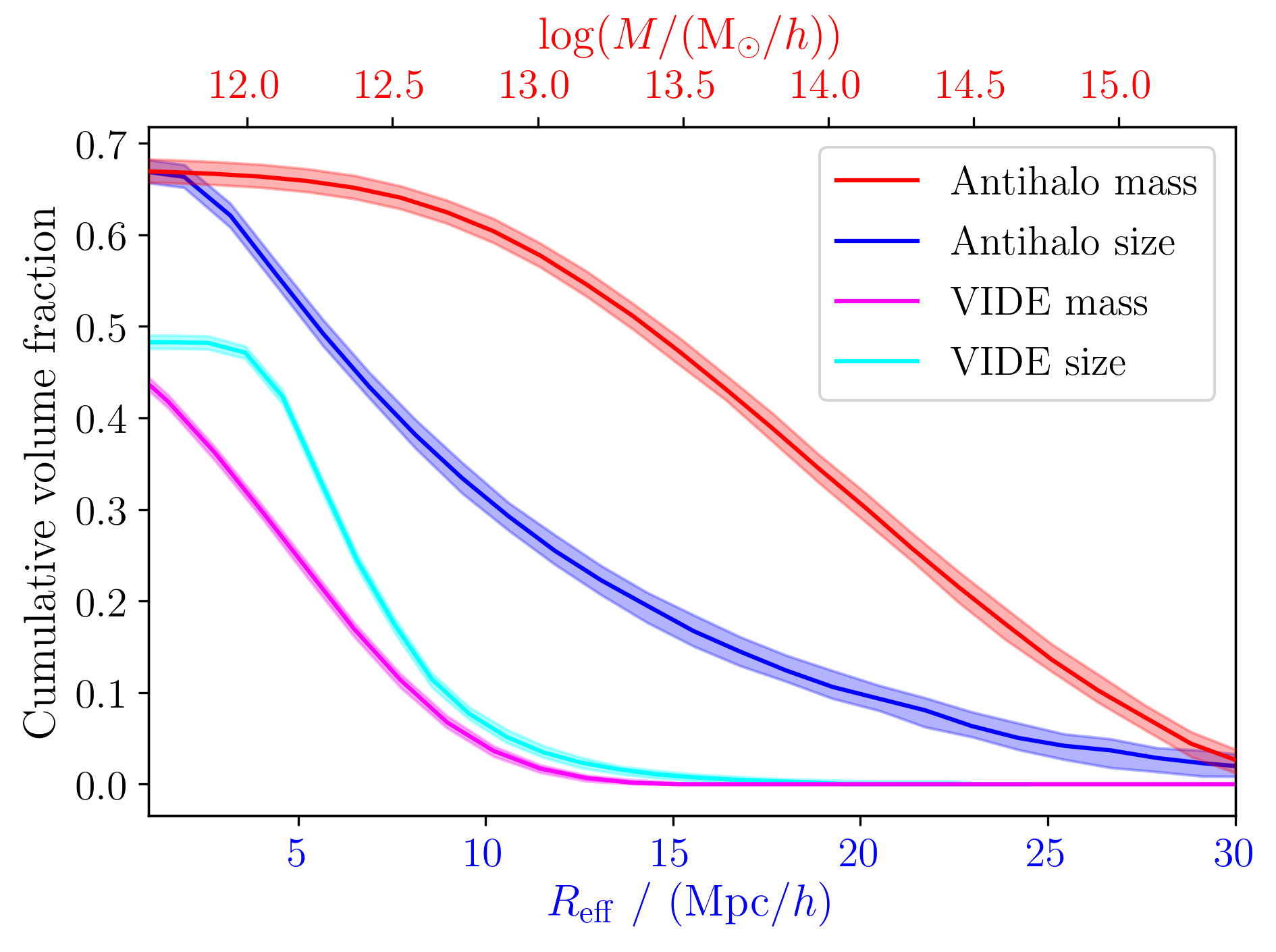}
  \caption{The cumulative volume filling fraction of voids as a function of either size (blue and cyan) or mass (red and magenta).}
  \label{fig:vol_frac}
\end{figure}

Fig.~\ref{fig:vol_frac} shows the cumulative volume filling fraction of antihalos and \texttt{VIDE} voids as a function of either $R_\text{eff}$ or $M$.
%along with the $68\%$ range among realisations as before.
Antihalos down to the minimum mass/size that we probe occupy $\sim$70\% of the zoom-in region, while those with $M>10^{15} \: \mathrm{M_\odot}/h$ alone already occupy $\sim$10\%. In contrast, even the full set of \texttt{VIDE} voids occupies $<50\%$ of the available volume.

%\red{Figs.~\ref{fig:axes}-\ref{fig:stacked_density} use step number \texttt{8740} of the \texttt{BORG} chain, but all realisations produce consistent results.}

% Finally, in Fig.~\ref{fig:AM} we compare the specific angular momenta of voids and halos: as would be expected, voids have much larger angular momentum because they are more extended. The angular momenta of voids in \cs\ and halos in \csi\ are only weakly correlated, indicating that they are predomantly generated during the evolution of the objects (different for void and halo regions), as opposed to being encoded in the initial conditions.

Although only included for three realisations in the public release, we also compute the antihalos' specific angular momenta around the VWB. This is typically in the range $1-10$ Mpc$^2$/s, around $100\times$ larger than for the halos in \csi.

% \begin{figure*}
%   \subfigure[]
%   {
%     \includegraphics[width=0.48\textwidth]{Figures/VSF_comb_avg}
%     \label{fig:VSF_diff}
%   }
%   \subfigure[]
%   {
%     \includegraphics[width=0.48\textwidth]{Figures/VSF_cum_comb_avg}
%     \label{fig:VSF_cum}
%   }
%   \caption{CAP}
%   \label{fig:VSF}
% \end{figure*}

% \begin{figure}
%   \centering
%   \includegraphics[width=0.5\textwidth]{Figures/M-R_7444}
%   \caption{CAP}
%   \label{fig:M-R}
% \end{figure}

%\begin{figure}
%  \centering
%  \includegraphics[width=0.5\textwidth]{Figures/rho_comp_8740}
%  \caption{The distributions of central and average overdensities of antihalos in two mass bins. The central density is defined as that of the sphere around the VWB enclosing 64 particles, while the average density includes all particles belonging to the void.}
%  \label{fig:overdensity}
%\end{figure}

% \begin{figure}
%   \centering
%   \includegraphics[width=0.5\textwidth]{Figures/Old7/AM_vel_comp}
%   \caption{CAP}
%   \label{fig:AM}
% \end{figure}

\subsection{Stacked density profiles}
\label{sec:dens}

\begin{figure}
  \centering
  \includegraphics[width=0.5\textwidth]{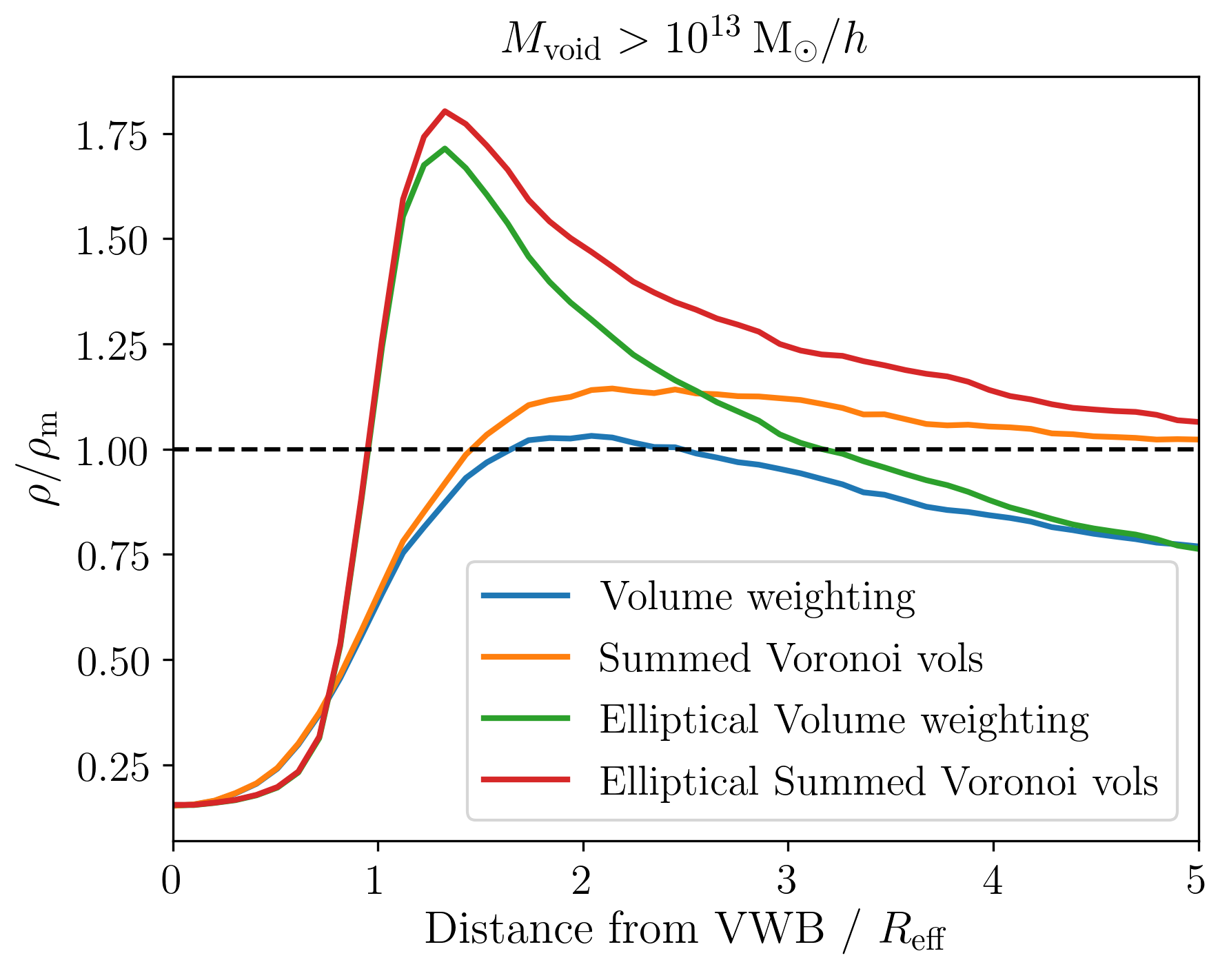}
  \caption{Stacked antihalo density profiles as a function of distance from the VWB in units of $R_\text{eff}$. We consider spherical and ellipsoidal averages, and for each we consider two methods for calculating the density (Eqs.~\ref{eq:dens2} and~\ref{eq:dens3}).}
  \label{fig:stacked_density}
\end{figure}

Fig.~\ref{fig:stacked_density} shows the stacked density profiles of antihalos with $M>10^{13} \: \mathrm{M_\odot}/h$ using the four methods described in Sec.~\ref{sec:dens}. All methods agree that the central regions are strongly underdense, with $\rho<0.2\:\rho_m$ on average. However, using an elliptical rather than spherical averaging results in a larger central core as well as a significantly sharper rise in density near $\reff$. This is because the elliptical method more accurately captures the 3D structure of the voids, while the spherical method washes out their features by combining regions that may be a very different number of principal axis lengths from the VWB. This also blurs out the ``ridge'' region at $\sim1-3\:\reff$ caused by surrounding filaments and walls.

The other important difference is that using the volumes of the shells rather than the Voronoi volumes of the particles themselves (Eq.~\ref{eq:dens2}, blue and green lines) produces a marked decline in the density profiles at $R \gtrsim 2\:\reff$. As already noted in \citet{Nadathur}, this is an artifact caused by the fact that the shells can extend beyond the region in which particles are found at large $R$, erroneously giving the impression that the density is falling below the cosmic mean. Given these significant differences, we recommend that future analyses account for the elliptical structure of voids
%(at least when defined as antihalos)
during stacking, and normalise densities by the Voronoi rather than shell volumes.

\subsection{Correlations and clustering}
\label{sec:corr}

% ACFs: antihalo-antihalo, VIDE-VIDE and halo-halo
% XCFs: antihalo-VIDE, antihalo-halo, VIDE-halo
% filtering ZOBOV voids on r_eff vs on mass

\begin{figure}
  \centering
  \includegraphics[width=0.5\textwidth]{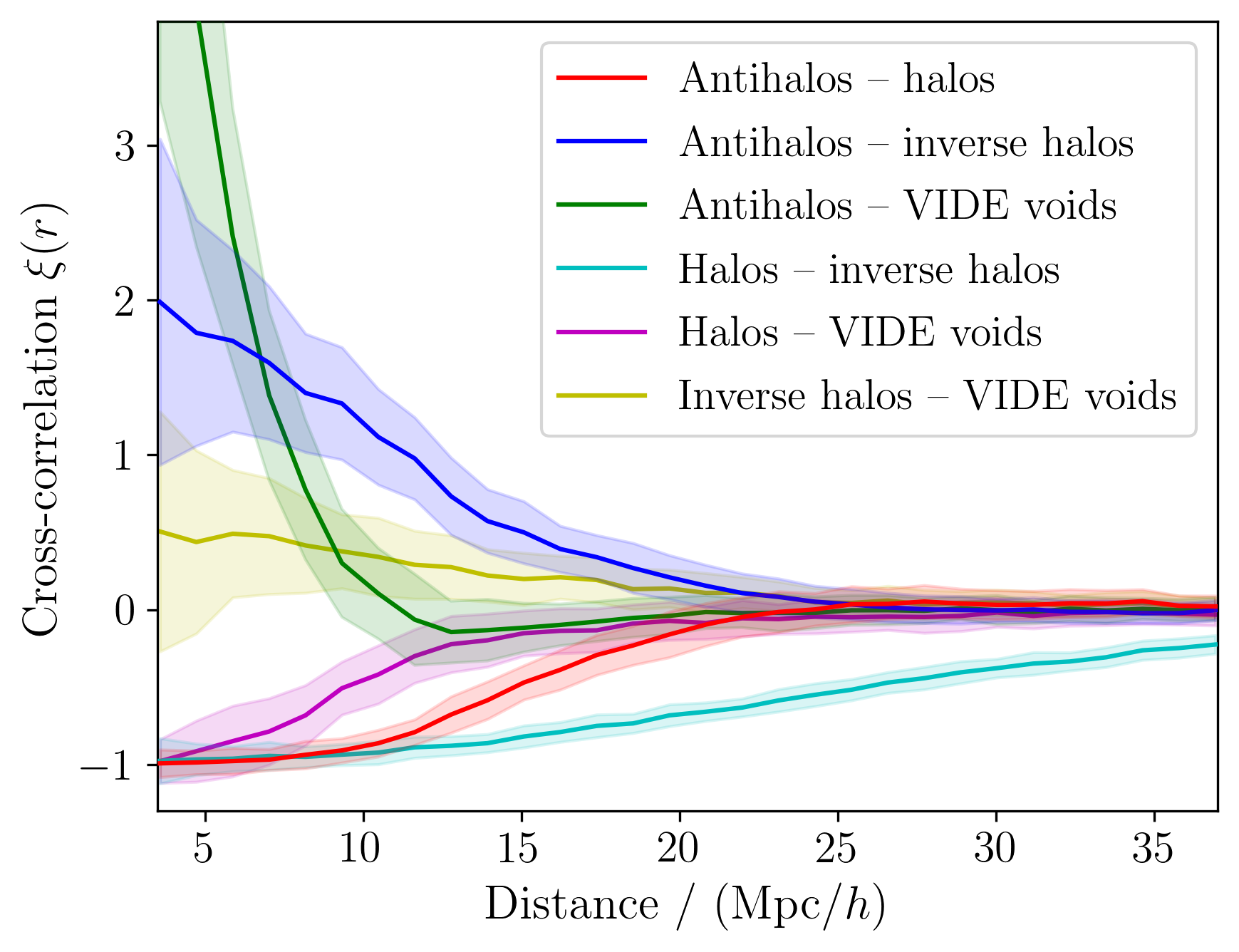}
  \caption{Cross-correlation functions between antihalos, \texttt{VIDE} voids, halos in \cs\ and halos in \csi\ (``inverse halos''), requiring $M > 10^{14} \: \mathrm{M_\odot}/h$ for antihalos, halos and inverse halos, and $\reff > 8.2 \: \text{Mpc}/h$ for \texttt{VIDE} voids.}
  \label{fig:corr}
\end{figure}

Fig.~\ref{fig:corr} shows the cross-correlation functions of the antihalos, \texttt{VIDE} voids,
%(with sub-sampling density 0.2 (Mpc/$h$)$^{-3}$)
and the halos in \cs\ and \csi\ (``inverse halos'') as a function of scale over the zoom-in region. We keep halos, antihalos and inverse halos with $M>10^{14} \: \mathrm{M_\odot}/h$, and \texttt{VIDE} voids with $\reff>8.2$ Mpc$/h$ (chosen such that the number of \texttt{VIDE} voids is similar to the number of antihalos). The clear positive correlation between the two void definitions and the inverse halos, and anticorrelation with the halos, affirms the expected antihalo behaviour. Applying the same mass cut to the \texttt{VIDE} catalogue 
%as to the antihalos
does not, however, yield the same results. Due to the weak correlation between $M$ and $\reff$ for \texttt{VIDE}, requiring $M > 10^{14} \: \mathrm{M_\odot}/h$ leaves many small voids in overdense regions producing a positive correlation with the halos. The almost monotonic increase of mass with radius for antihalos makes their spatial distribution much more robust to the cut employed. We have also calculated the autocorrelation functions, and find that while $M \gtrsim 10^{13} \: \mathrm{M_\odot}/h$ antihalos anticorrelate on all accessible scales, the \texttt{VIDE} voids only anticorrelate due to exclusion on scales comparable to their size. On larger scales, mass-selected samples can autocorrelate as strongly as halos. Correlation functions with any cuts may be easily explored using the public catalogues. Note that voids small compared to the $2.7$ Mpc/$h$ resolution of the \texttt{BORG} ICs are poorly constrained, with large variance between different \csi\ realisations, so should be removed in studies where accurate positions are required.

%\mh{Void exclusion in VIDE $\rightarrow$ anti-correlation only on scales $\sim$ void size.}

\subsection{Effect of the \texttt{BORG} ICs}
\label{sec:borg_ics}

To investigate the impact of the \texttt{BORG} constraints we have run a pair of simulations identical to \cs\ but with random ICs. We find the antihalo results of Figs.~\ref{fig:axes}, \ref{fig:stacked_density} and \ref{fig:corr} to be consistent with \cs, although we find somewhat more large halos and antihalos, and correspondingly fewer small halos and antihalos, compared to the unconstrained case. A similar trend is seen in the antihalo mass function, which exceeds that of \citet{Warren} for $\log(M/M_\odot)\gtrsim14.35$ and lies $\sim$0.1 dex below it at lower masses. This may indicate $\sim$3$\sigma$ above average structure in the 2M++ volume compared to the cosmic mean or may be an artifact of the \texttt{BORG} algorithm or \texttt{PHEW} halofinder, and will be the subject of future work. These trends cannot be accounted for by altering $\Omega_m$ in the \citeauthor{Warren} curve to match the average simulation density.

\section{Discussion and Conclusion}
\label{sec:disc_conc}

% //Try with different halofinders
% //Future work in terms of applications
% //Anti-subhalos?
% Anything like resolution that would improve the quality of the voids?
% Making the data publicly available

We run a suite of simulations constrained to match the local Universe by using ICs inferred from the \texttt{BORG} algorithm applied to the 2M++ galaxy catalogue, and a corresponding suite with the ICs inverted. This enables us to apply a new voidfinder out to 155 Mpc/$h$, in which voids are defined by dark matter particles belonging to halos in the inverted simulation. Running 101 inverted and uninverted simulations with ICs selected from across the \texttt{BORG} posterior allows us to propagate
%the uncertainties in \texttt{BORG}
its uncertainty into our void catalogues, including the impact of density modes below the \texttt{BORG} grid. A key advantage of this voidfinder over traditional watershed methods is that the voids it identifies are linear to smaller scales and lower redshifts.
%, and may be described by the excursion set formalism well established for halos in the inverted simulation \citep{Press_Schechter, Sheth_Tormen}.
In addition, the higher tracer density associated with the use of dark matter particles allows smaller voids to be identified and all voids to be characterised in more detail. We calculate a number of standard diagnostics of antihalos, finding a more extended size function, greater volume filling fraction, greater ellipticity and more stable correlation behaviour compared to \texttt{VIDE} voids. We make our antihalo and \texttt{VIDE} catalogues publicly available.
%-- mass and size functions, volume filling fraction, ellipticity, central and average density, specific angular momentum and stacked density profiles --
%and make the catalogues publicly available for specific applications.

Our work leaves a number of important questions for the voids-as-antihalos method. The most important parameters of the voidfinder are those of the halofinder applied to the inverted simulation. While the \texttt{PHEW} algorithm that we use has been shown to reproduce standard statistics with the default settings \citep{PHEW}, it will be important to investigate how varying these impacts the void population. It would also be useful to repeat the analysis using alternative algorithms such as friends-of-friends \citep{FOF}, \texttt{HOP} \citep{HOP} or a phase-space halofinder.
%such as \texttt{ROCKSTAR} \citep{Rockstar}.
Note that the halofinder parameters do not contribute systematic uncertainty to the void properties but instead lead to different void definitions; the optimal parameters will be those that maximise sensitivity to whatever one wishes to test in a concrete application. The uncertainty in the void properties derives from that in the ICs (which our use of effectively the full \texttt{BORG} posterior allows us to marginalise over) as well as any approximations made during the simulation. One could also broaden the antihalo definition to include subhalos in the inverted simulation: it would be worth seeing if anti-subhalos have systematically different properties to anti-main-halos, and if they carry additional astrophysical or cosmological information. It would also be interesting to contrast their formation histories with those of halos, both readily available within our framework, and compare to voids found directly in the galaxy field.
%(e.g. from the 2M++ catalogue).
% The quality of the voids is of course dependent on the specifications of the constrained simulations, so users must ensure that these meet their requirements.

Our void catalogues may find a range of applications, including studying secondary anisotropies of the Cosmic Microwave Background, characterising cosmic gas pressure, growth rate and star formation rate at low density, probing cosmological parameters and initial conditions,
%such as primordial non-Gaussianity,
and searching for modified gravity (see \citealt{white_paper} and references therein). We will explore some of these in future work. Note that while the \texttt{BORG} inference on which our analysis is based assumes $\Lambda$CDM, it is straightforward to incorporate additional physics within it and effort is underway to do so. This will enable direct comparison of the local voids produced under various scenarios. Similar catalogues could already be made in the SDSS-III/BOSS volume using the corresponding \texttt{BORG} reconstruction \citep{BORG_SDSS}, and application of the algorithm to upcoming datasets from Euclid, the Roman Space Telescope or the Vera Rubin Observatory will greatly increase the accessible redshift range. Given the advantages of the antihalo definition,
%shown here
we believe that methods along these lines will be integral to future void science.

% - Improve on existing detections of the tSZ effect in voids, and use them to make a measurement of the mean cosmic gas pressure (as in https://arxiv.org/abs/1709.01489).
% - Attempt a detection of the kSZ effect via pairwise velocity correlations using void-void and void-galaxy pairs.
% - In general, stacking CMB observables on voids (CMB map for ISW, CMB lensing, tSZ, CIB) to study different quantities (growth, density, pressure, SFR) in low densities.
% - Compare basic void properties (depth distribution, profile, abundance) with "geometric" (as opposed to dynamical) void finders.
% - Study void profile universality and void "mass function" universality (where "mass function" here would mean abundance of voids as a function of the mass of the halo in the reversed sim). I guess this is not something you'd do with a void catalog from data, but from simulations though...
% - Study sensitivity to initial conditions (i.e. primordial non-Gaussianity) of void abundance and void correlation function (again, this would be probably from sims...).

% \section{Conclusions}
% \label{sec:conc}

\section{Data availability}

%Our antihalo and \texttt{VIDE} catalogues are available at \url{https://zenodo.org/record/5503610}. Other data may be shared upon request to the corresponding author.
Our antihalo and \texttt{VIDE} catalogues are available \href{https://zenodo.org/record/5503610}{here}. Other data may be shared on request to the corresponding author.

\section*{Acknowledgements}

%We are indebted to David Alonso, Roberto Gonzalez, Nico Hamaus, Mladen Ivkovic, Jens Jasche, Guilhem Lavaux, Chris Rycroft and Romain Teyssier for input and discussions.
% We thank David Alonso, Chris Rycroft, Roberto Gonzalez, Mladen Ivkovic, Romain Teyssier ... for useful discussions.
% We are very grateful to
% We are indebted to

We are indebted to D. Alonso, R. Gonzalez, N. Hamaus, M. Ivkovic, J. Jasche, G. Lavaux, C. Rycroft and R. Teyssier for input and discussions. HD is supported by St John's College, Oxford. 
%, and acknowledges financial support from ERC Grant No. 693024 and the Beecroft Trust.
This work was done within the \href{https://www.aquila-consortium.org/}{Aquila Consortium}.

This work used the DiRAC Complexity and DiRAC@Durham facilities, operated by the University of Leicester IT Services and Institute for Computational Cosmology, which form part of the STFC DiRAC HPC Facility (www.dirac.ac.uk). This equipment is funded by BIS National E-Infrastructure capital grants ST/K000373/1, ST/P002293/1, ST/R002371/1 and ST/S002502/1, STFC DiRAC Operations grant ST/K0003259/1, and Durham University and STFC operations grant ST/R000832/1. DiRAC is part of the National E-Infrastructure.

%This work was done within the Aquila Consortium (\url{https://www.aquila-consortium.org/}).

% This work used the DiRAC@Durham facility managed by the Institute for Computational Cosmology on behalf of the STFC DiRAC HPC Facility (www.dirac.ac.uk). The equipment was funded by BEIS capital funding via STFC capital grants ST/P002293/1, ST/R002371/1 and ST/S002502/1, Durham University and STFC operations grant ST/R000832/1. DiRAC is part of the National e-Infrastructure.

\bibliographystyle{mnras}
\bibliography{references}

\begin{thebibliography}{}
\makeatletter
\relax
\def\mn@urlcharsother{\let\do\@makeother \do\$\do\&\do\#\do\^\do\_\do\%\do\~}
\def\mn@doi{\begingroup\mn@urlcharsother \@ifnextchar [ {\mn@doi@}
  {\mn@doi@[]}}
\def\mn@doi@[#1]#2{\def\@tempa{#1}\ifx\@tempa\@empty \href
  {http://dx.doi.org/#2} {doi:#2}\else \href {http://dx.doi.org/#2} {#1}\fi
  \endgroup}
\def\mn@eprint#1#2{\mn@eprint@#1:#2::\@nil}
\def\mn@eprint@arXiv#1{\href {http://arxiv.org/abs/#1} {{\tt arXiv:#1}}}
\def\mn@eprint@dblp#1{\href {http://dblp.uni-trier.de/rec/bibtex/#1.xml}
  {dblp:#1}}
\def\mn@eprint@#1:#2:#3:#4\@nil{\def\@tempa {#1}\def\@tempb {#2}\def\@tempc
  {#3}\ifx \@tempc \@empty \let \@tempc \@tempb \let \@tempb \@tempa \fi \ifx
  \@tempb \@empty \def\@tempb {arXiv}\fi \@ifundefined
  {mn@eprint@\@tempb}{\@tempb:\@tempc}{\expandafter \expandafter \csname
  mn@eprint@\@tempb\endcsname \expandafter{\@tempc}}}

\bibitem[\protect\citeauthoryear{{Bartlett}, {Desmond}  \&
  {Ferreira}}{{Bartlett} et~al.}{2021}]{Bartlett}
{Bartlett} D.~J.,  {Desmond} H.,   {Ferreira} P.~G.,  2021, \mn@doi [\prd]
  {10.1103/PhysRevD.103.023523}, \href
  {https://ui.adsabs.harvard.edu/abs/2021PhRvD.103b3523B} {103, 023523}

\bibitem[\protect\citeauthoryear{{Bleuler}, {Teyssier}, {Carassou}  \&
  {Martizzi}}{{Bleuler} et~al.}{2015}]{PHEW}
{Bleuler} A.,  {Teyssier} R.,  {Carassou} S.,   {Martizzi} D.,  2015, \mn@doi
  [Computational Astrophysics and Cosmology] {10.1186/s40668-015-0009-7}, \href
  {https://ui.adsabs.harvard.edu/abs/2015ComAC...2....5B} {2, 5}

\bibitem[\protect\citeauthoryear{{Davis}, {Efstathiou}, {Frenk}  \&
  {White}}{{Davis} et~al.}{1985}]{FOF}
{Davis} M.,  {Efstathiou} G.,  {Frenk} C.~S.,   {White} S.~D.~M.,  1985,
  \mn@doi [\apj] {10.1086/163168}, \href
  {https://ui.adsabs.harvard.edu/abs/1985ApJ...292..371D} {292, 371}

\bibitem[\protect\citeauthoryear{Desmond, Hutt, Devriendt  \& Slyz}{Desmond
  et~al.}{2021}]{zenodo}
Desmond H.,  Hutt M.~L.,  Devriendt J.,   Slyz A.,  2021,
  \mn@doi{10.5281/zenodo.5503610}

\bibitem[\protect\citeauthoryear{{Eisenstein} \& {Hut}}{{Eisenstein} \&
  {Hut}}{1998}]{HOP}
{Eisenstein} D.~J.,  {Hut} P.,  1998, \mn@doi [\apj] {10.1086/305535}, \href
  {https://ui.adsabs.harvard.edu/abs/1998ApJ...498..137E} {498, 137}

\bibitem[\protect\citeauthoryear{{Jasche} \& {Lavaux}}{{Jasche} \&
  {Lavaux}}{2019}]{BORG_PM}
{Jasche} J.,  {Lavaux} G.,  2019, \mn@doi [\aap] {10.1051/0004-6361/201833710},
  \href {https://ui.adsabs.harvard.edu/abs/2019A&A...625A..64J} {625, A64}

\bibitem[\protect\citeauthoryear{{Jasche} \& {Wandelt}}{{Jasche} \&
  {Wandelt}}{2012}]{BORG_1}
{Jasche} J.,  {Wandelt} B.~D.,  2012, \mn@doi [\mnras]
  {10.1111/j.1365-2966.2012.21423.x}, \href
  {https://ui.adsabs.harvard.edu/abs/2012MNRAS.425.1042J} {425, 1042}

\bibitem[\protect\citeauthoryear{{Jasche} \& {Wandelt}}{{Jasche} \&
  {Wandelt}}{2013}]{BORG_2}
{Jasche} J.,  {Wandelt} B.~D.,  2013, \mn@doi [\mnras] {10.1093/mnras/stt449},
  \href {https://ui.adsabs.harvard.edu/abs/2013MNRAS.432..894J} {432, 894}

\bibitem[\protect\citeauthoryear{{Lavaux} \& {Hudson}}{{Lavaux} \&
  {Hudson}}{2011}]{2Mpp}
{Lavaux} G.,  {Hudson} M.~J.,  2011, \mn@doi [\mnras]
  {10.1111/j.1365-2966.2011.19233.x}, \href
  {https://ui.adsabs.harvard.edu/abs/2011MNRAS.416.2840L} {416, 2840}

\bibitem[\protect\citeauthoryear{{Lavaux} \& {Jasche}}{{Lavaux} \&
  {Jasche}}{2016}]{BORG_2Mpp}
{Lavaux} G.,  {Jasche} J.,  2016, \mn@doi [\mnras] {10.1093/mnras/stv2499},
  \href {https://ui.adsabs.harvard.edu/abs/2016MNRAS.455.3169L} {455, 3169}

\bibitem[\protect\citeauthoryear{{Lavaux}, {Jasche}  \& {Leclercq}}{{Lavaux}
  et~al.}{2019}]{BORG_SDSS}
{Lavaux} G.,  {Jasche} J.,   {Leclercq} F.,  2019, arXiv e-prints, \href
  {https://ui.adsabs.harvard.edu/abs/2019arXiv190906396L} {p. arXiv:1909.06396}

\bibitem[\protect\citeauthoryear{{Leclercq}, {Jasche}, {Sutter}, {Hamaus}  \&
  {Wandelt}}{{Leclercq} et~al.}{2015}]{Florent_1}
{Leclercq} F.,  {Jasche} J.,  {Sutter} P.~M.,  {Hamaus} N.,   {Wandelt} B.,
  2015, \mn@doi [\jcap] {10.1088/1475-7516/2015/03/047}, \href
  {https://ui.adsabs.harvard.edu/abs/2015JCAP...03..047L} {2015, 047}

\bibitem[\protect\citeauthoryear{{Nadathur} \& {Hotchkiss}}{{Nadathur} \&
  {Hotchkiss}}{2015}]{Nadathur_Hotchkiss}
{Nadathur} S.,  {Hotchkiss} S.,  2015, \mn@doi [\mnras]
  {10.1093/mnras/stv2131}, \href
  {https://ui.adsabs.harvard.edu/abs/2015MNRAS.454.2228N} {454, 2228}

\bibitem[\protect\citeauthoryear{{Nadathur}, {Hotchkiss}, {Diego}, {Iliev},
  {Gottl{\"o}ber}, {Watson}  \& {Yepes}}{{Nadathur} et~al.}{2015}]{Nadathur}
{Nadathur} S.,  {Hotchkiss} S.,  {Diego} J.~M.,  {Iliev} I.~T.,
  {Gottl{\"o}ber} S.,  {Watson} W.~A.,   {Yepes} G.,  2015, \mn@doi [\mnras]
  {10.1093/mnras/stv513}, \href
  {https://ui.adsabs.harvard.edu/abs/2015MNRAS.449.3997N} {449, 3997}

\bibitem[\protect\citeauthoryear{{Neyrinck}}{{Neyrinck}}{2008}]{Zobov}
{Neyrinck} M.~C.,  2008, \mn@doi [\mnras] {10.1111/j.1365-2966.2008.13180.x},
  \href {https://ui.adsabs.harvard.edu/abs/2008MNRAS.386.2101N} {386, 2101}

\bibitem[\protect\citeauthoryear{{Pisani} et~al.,}{{Pisani}
  et~al.}{2019}]{white_paper}
{Pisani} A.,  et~al., 2019, \baas, \href
  {https://ui.adsabs.harvard.edu/abs/2019BAAS...51c..40P} {51, 40}

\bibitem[\protect\citeauthoryear{{Pontzen}, {Slosar}, {Roth}  \&
  {Peiris}}{{Pontzen} et~al.}{2016}]{Pontzen_1}
{Pontzen} A.,  {Slosar} A.,  {Roth} N.,   {Peiris} H.~V.,  2016, \mn@doi [\prd]
  {10.1103/PhysRevD.93.103519}, \href
  {https://ui.adsabs.harvard.edu/abs/2016PhRvD..93j3519P} {93, 103519}

\bibitem[\protect\citeauthoryear{{Rycroft}}{{Rycroft}}{2009}]{Voro}
{Rycroft} C.~H.,  2009, \mn@doi [Chaos] {10.1063/1.3215722}, \href
  {https://ui.adsabs.harvard.edu/abs/2009Chaos..19d1111R} {19, 041111}

\bibitem[\protect\citeauthoryear{{Shim}, {Park}, {Kim}  \& {Hwang}}{{Shim}
  et~al.}{2021}]{AntiClusters}
{Shim} J.,  {Park} C.,  {Kim} J.,   {Hwang} H.~S.,  2021, \mn@doi [\apj]
  {10.3847/1538-4357/abd0f6}, \href
  {https://ui.adsabs.harvard.edu/abs/2021ApJ...908..211S} {908, 211}

\bibitem[\protect\citeauthoryear{{Sinha} \& {Garrison}}{{Sinha} \&
  {Garrison}}{2020}]{corrfunc}
{Sinha} M.,  {Garrison} L.~H.,  2020, \mn@doi [\mnras] {10.1093/mnras/stz3157},
  \href {https://ui.adsabs.harvard.edu/abs/2020MNRAS.491.3022S} {491, 3022}

\bibitem[\protect\citeauthoryear{{Stopyra}, {Peiris}  \& {Pontzen}}{{Stopyra}
  et~al.}{2021}]{Pontzen}
{Stopyra} S.,  {Peiris} H.~V.,   {Pontzen} A.,  2021, \mn@doi [\mnras]
  {10.1093/mnras/staa3587}, \href
  {https://ui.adsabs.harvard.edu/abs/2021MNRAS.500.4173S} {500, 4173}

\bibitem[\protect\citeauthoryear{{Sutter} et~al.,}{{Sutter}
  et~al.}{2015}]{VIDE}
{Sutter} P.~M.,  et~al., 2015, \mn@doi [Astronomy and Computing]
  {10.1016/j.ascom.2014.10.002}, \href
  {https://ui.adsabs.harvard.edu/abs/2015A&C.....9....1S} {9, 1}

\bibitem[\protect\citeauthoryear{{Teyssier}}{{Teyssier}}{2002}]{Ramses}
{Teyssier} R.,  2002, \mn@doi [\aap] {10.1051/0004-6361:20011817}, \href
  {https://ui.adsabs.harvard.edu/abs/2002A&A...385..337T} {385, 337}

\bibitem[\protect\citeauthoryear{{Warren}, {Abazajian}, {Holz}  \&
  {Teodoro}}{{Warren} et~al.}{2006}]{Warren}
{Warren} M.~S.,  {Abazajian} K.,  {Holz} D.~E.,   {Teodoro} L.,  2006, \mn@doi
  [\apj] {10.1086/504962}, \href
  {https://ui.adsabs.harvard.edu/abs/2006ApJ...646..881W} {646, 881}

\makeatother
\end{thebibliography}

%\bsp
\label{lastpage}
\end{document}